\documentclass{ifacconf}
\usepackage{bbm}
\usepackage{natbib}            
\usepackage{graphicx}          
\RequirePackage{xspace}
\RequirePackage{amsmath, amssymb, textcomp}
\usepackage{amsmath} 
\usepackage{amssymb}  
\usepackage{graphicx,url}
\usepackage{bbold,dsfont,color}
\newcommand{\T}{\mathsf{T}}
\renewcommand{\b}{{ b}}

\newcommand{\xx}{{\bf x}}

\DeclareMathOperator*\argmin{arg\,min}

\newcommand{\DD}{{\bf D}}

\newcommand{\bb}{{\bf b}}
\renewcommand{\aa}{{\bf a}}

\newcommand{\uu}{{\bf  u}}

\newcommand{\X}{{\bf X}}

\newcommand{\yy}{{\bf y}}
\newcommand{\y}{{ y}}
\renewcommand{\a}{a}
\newcommand{\eg}{\textit{e.g.,~}}
\newcommand{\ie}{\textit{i.e.,~}}
\def\subjto{{\mbox{subj. to}}}

\renewcommand{\Re}{{\mathbb{R}}}
\usepackage{tikz}
\usetikzlibrary{arrows}
\usetikzlibrary{decorations.markings}

\begin{document}

\begin{frontmatter}

\title{Blind Identification via Lifting\thanksref{footnoteinfo}} 

\thanks[footnoteinfo]{The work presented is supported by the NSF
Graduate Research Fellowship under grant DGE 1106400, NSF
CPS:Large:ActionWebs award number 0931843, TRUST (Team for Research in
Ubiquitous Secure Technology) which receives support from NSF (award
number CCF-0424422), and FORCES (Foundations Of Resilient
CybEr-physical Systems), the Swedish Research
  Council in the Linnaeus center CADICS, the European Research Council
   under the advanced grant LEARN, contract 267381, a postdoctoral grant from the Sweden-America
   Foundation, donated by ASEA's Fellowship Fund, and  by a postdoctoral
   grant from the Swedish Research Council.}

\author[Ber,Liu]{Henrik Ohlsson} 
\author[Ber]{Lillian Ratliff} 
\author[Ber]{Roy Dong}
\author[Ber]{S. Shankar Sastry}

\address[Ber]{Department of Electrical Engineering and Computer
  Sciences, University of California, Berkeley, CA, USA (e-mail:
  ohlsson@eecs.berkeley.edu).}                                              
\address[Liu]{Division of Automatic Control, Department of Electrical Engineering, Link\"oping University, Sweden.}


\begin{abstract}                          
Blind system identification  is known to be an ill-posed problem and
without further assumptions, no unique solution is at hand. In this contribution, we are concerned with the task of identifying an ARX model
from only output measurements. 
We phrase this as a constrained
rank minimization problem and present a relaxed convex formulation to approximate
its solution. To make the problem well posed we assume that the sought
input lies in some known linear subspace.
\end{abstract}

\end{frontmatter}

\section{Introduction}

Consider an auto-regressive exogenous input (ARX) model   
\begin{align}\nonumber
y(t)-&\a_1 y(t-1)  
- \cdots - \a_{n_a} y(t-n_a) \\=&
\b_1 u(t-n_k - 1) 
+\cdots +\b_{n_b} u(t-n_k - n_b) 
\end{align} with input $u \in \Re$ and output $y \in \Re$.
Estimation of this type of model is one of  the most common tasks in
system identification and a very well studied problem, see for instance \citet{Ljung:99}. 
The common
setting is that $\{(y(t),u(t))\}_{t=1}^N$  is given and the summed
residuals{\small
\begin{equation*}
\sum_{t=n}^N \left (y(t) - \sum_{k_1=1}^{n_b}
\b_{k_1} u(t-k_1-n_k) - \sum_{k_2=1}^{n_a}
\a_{k_2} y(t-k_2) \right )^2
\end{equation*}}%
where $n=\max( n_a, n_k+n_b)+1$, is minimized to obtain an estimate for $\a_1,\dots, \a_{n_a},\b_1,\dots, \b_{n_b}$. This estimate is often
referred to as the \textit{least squares} (LS) estimate.

In this paper we study the more complicated problem of
estimating an ARX model from solely outputs $\{y(t)\}_{t=1}^N$. This is
an ill-posed problem and it is easy to see that under no further
assumptions, it would be impossible to uniquely determine
$\a_1,\dots, \a_{n_a},\b_1,\dots, \b_{n_b}$. We will here therefore study this problem under the assumption
that the stacked inputs belong to some known subspace. The input
could for example be:
\begin{itemize}
\item known to change only at a set of discrete times
due to a discrete controller or  
\item known to be band-limited and therefore well represented by the projection
  on the first discrete Fourier transform basis vectors.  
\end{itemize}
It should be noticed that this assumption 
is not enough to uniquely determine the input or the ARX
model. Specifically, we will not be able to decide the input or the
ARX coefficients $\b_1,\dots, \b_{n_b}$  more than up to a
multiplicative scalar. It should be stressed that this is not a
limitation of the method that we propose but an inherent
limitation of the system identification problem since the sought quantities
always appear as products. To uniquely determine the input and the
ARX coefficients $\b_1,\dots, \b_{n_b}$, further knowledge is needed.

The main contribution of the paper is a novel method for ARX model
identification from only output measurements. The method takes the
form of a convex optimization problem and gives a computationally  flexible framework
for handling different types of measurement noises, constraints, \textit{etc}.  


\section{Background}
\label{sec:background}

Blind system identifican (BSI) has a broad application area and has
been applied in fields such as  data communications, speech
recognition and seismic signal processing, see for instance
\cite{Abed97}. Common for the type of modeling problems  that BSI has been
applied to is that the input is difficult, costly or impossible to
measure. In for example  exploration seismology, the physical
properties of the earth are explored by studying the response of
 an excitation (often a charge of dynamite). The excitation is often
 difficult to measure and the modeling problem therefore a BSI problem, see \eg \cite{Zerva199947}.

Many methods have been proposed to solve the BSI problem throughout
the years. We give a short overview here but refer the interested
reader to  \cite{Abed97,Hua02}, for a more extensive and complete review. 

The maximum likelihood (ML) approach to BSI aims at finding the ML
estimate of the model and input. The resulting non-convex optimization
problem is often treated by alternating between optimizing with respect to
the input and the system model.  See \cite{meraim:EUSIPCO:1994}. The
channel subspace (CS) methods to BSI  indirectly determine the sought
finite impulse response (FIR)  model by estimating the nullspace of
the Sylvester matrix associated with the FIR model to be identified. This is
done by an eigen decomposition of a matrix derived from the
outputs. See for instance \cite{Abed-Meraim:06}.
The method proposed in \cite{Zerva00} works under the assumption that
two or more output series are available and that these were generated
by the same input. See also \cite{Van13}. The methods proposed in \cite{Sato75,Tong91}
assume that the input consists of  independent and identically
(iid)  distributed random variables and considers the autocorrelation of the
output to decide a FIR model and the unknown
input. 

A number of approaches consider the blind identification problem of
Hammerstein systems under the assumption that the input is piecewise
constant. \cite{Sun99,BaiLD02,Bai02,Wang07,Wang09,Wang10}. Our approach assumes that the input belongs to
some known subspace.  Piecewise constant signal can be represented
using the subspace assumption used here. However, we note that we
are not restricted to piecewise constant signals, and our approach is
significantly different. Also, we consider the blind identification of
ARX models while the blind identification problem of
Hammerstein systems is considered in
\cite{Sun99,BaiLD02,Bai02,Wang07,Wang09,Wang10}.

The related problem of blind deconvolution have
been studied in a number of contributions. In particular,  see the
very interesting paper by
\cite{Ahmed:12} for a solution where the signals to be recovered are
assumed to be in some known subspaces. The development presented in
\cite{Ahmed:12} has similarities to the approach presented in this
paper and was done in parellel to our work. Note that only FIR models
are discussed in \cite{Ahmed:12}  and that the analysis does not apply.

\section{Problem Formulation}
Given the sequence of outputs $\{y(t)\}_{t=1}^N \in \Re$, 
find an estimate for  $\a_1,\dots,
\a_{n_a},\b_1,\dots, \b_{n_b} \in \Re$ and  $u(t) \in \Re,t=1,\dots,N,$ such that  
\begin{align*}\nonumber
y(t)-&\a_1 y(t-1) 
- \cdots - \a_{n_a} y(t-n_a) \\=&
\b_1 u(t-n_k - 1) 
+\cdots +\b_{n_b} u(t-n_k - n_b)+w(t),
\end{align*} for  $t=n,\dots,N$, where $n=\max( n_a, n_k+n_b)+1$, and $w(t),\,t=n,\dots,N$, 
some unknown zero mean noise.
We will for simplicity assume that $n_a, n_b, n_k,$ are
known. To make the problem well posed, we will seek an input in a given
subspace of $\Re^N$.

\section{Notation and Assumptions}

We will use $y$ to denote the output and $u$ the input. We will for
simplicity only consider \textit{single input single output} (SISO)
systems, however with some extra bookkeeping also MIMO systems could
be treated. We will assume
that $N$ measurements of $y$ are available and stack them in the vector
$\yy$, \ie \begin{align} \yy=\begin{bmatrix} \y(1) &\dots &
   \y(N)  \end{bmatrix}^\T.\end{align}  We also introduce $\uu$, $\eta$, $\aa$
and $\bb$
 as
\begin{align} 
\uu=&\begin{bmatrix} u(1) &\dots &
    u(N)  \end{bmatrix}^\T,\\
 {\eta}=&\begin{bmatrix} \eta(1) &\dots &
   \eta(N) \end{bmatrix}^\T,\\ 
\aa=&\begin{bmatrix} \a_1 &\dots &
    \a_{n_a}  \end{bmatrix}^\T,\\
\bb=&\begin{bmatrix} \b_1 &\dots &
    \b_{n_b}  \end{bmatrix}^\T.\end{align} We will use $\yy(i)$ to
denote the $i$th element of $\yy$. To pick out a subvector of $\yy$
consisting of the $i$th to the $j$th element we will use the notation
$\yy(i:j)$ and similarly for picking out a subvector of $\uu$, $\aa$
and $\bb$. To pick out a submatrix consisting of the $i$th to the
$j$th rows of $\X$ we use the notation $\X(i:j,:)$.
We will use normal font to represent scalars and bold for vectors and
matrices. 
 $\|\cdot \|_0$ is the zero (quasi) norm which returns the
number of nonzero elements of its argument and $\|\cdot\|_*$ the
nuclear norm returning the sum of the singular values.

We will assume that it is known that the sought input, $\uu$, lies in
some known subspace. We can hence write 
\begin{equation}\label{eq:assum}
\uu=\DD\xx
\end{equation}
for some known $N\times m$-matrix $\DD$ and an unknown vector $\xx \in
\Re^m$. It is assumed that $m\leq N$.
\section{Blind Identification via Lifting}


Consider the noise free setting where $w(t)=0, \,t=n,\dots,N.$
We can formulate the problem of finding an input 
and the ARX coefficients as the feasibility 
problem
\begin{subequations}\label{eq:probform}
\begin{align}\label{eq:probform1}
& \hspace*{-2.05cm}\text{find}_{\begin{array}{cc}u(t),\,t=1,\dots,N,\\ \a_1,\dots,
\a_{n_a}, \b_1,\dots, \b_{n_b} \end{array}}  \quad  
\\  \hspace{0cm} \nonumber
  \subjto   \quad y(t)-&\sum_{k_2=1}^{n_a}\a_{k_2} y(t-k_2)
  \\   \label{eq:probform2}  =& \sum_{k_1=1}^{n_b} 
\b_{k_1} u(t-n_k - k_1), \quad  t=n,\dots,N.
\end{align}\end{subequations}
Note that the 
  $\{
\a_k\}_{k=1}^{n_a}$, $\{ \b_k\}_{k=1}^{n_b}$,  
and  $ \{
u(t)\}_{t=1}^N $ are unknown. The problem is therefore non-convex. 

Introduce 
$\X = \xx \bb^\T \in \Re ^ {m    \times n_b}$ and note that
\eqref{eq:assum} gives that $\DD
\X=\DD \xx \bb^\T=\uu \bb^\T$.  Since $\uu \bb^\T$ contains all
products  $u(i) b_j,\, i=1,\dots, N, \, j=1,\dots, n_b$, the sum 
\begin{equation}\sum_{k_1=1}^{n_b} 
\b_{k_1} u(t-n_k - k_1)\end{equation}
can be realized by summing appropriate entries of $\DD \X$.
Problem \eqref{eq:probform}  can  now be reformulated as
\begin{subequations}\label{eq:probformm}
\begin{align}\label{eq:probformm1}
\text{find}_{\X,\aa}  \quad   &\\ \nonumber
 \subjto \quad  y(t) &-\sum_{k_2=1}^{n_a} \a_{k_2} y(t-k_2)
\\   \label{eq:probformm2} \quad & = \sum_{k_1=1}^{n_b} (\DD\X)(t-n_k-k_1,k_1),\; t=n,\dots,N, \\ \quad  rank(\X) &=1.
\end{align}\end{subequations} 
Note that we need to require that $rank(\X)=1$ to not lose the
possibility to decompose $\X$ as $ \X=\xx \bb^\T$.
The problem \eqref{eq:probformm} is equivalent to \eqref{eq:probform}  in the following
sense. Assume that \eqref{eq:probformm}, has a unique solution $\X^*$,
then  $\X^*$ must satisfy
$\X^*=\xx^* (\bb^*)^\T$, with $\xx^*$ and $\bb^*$ solving
\eqref{eq:probform}. Extracting the rank 1 component of $\X^*$, using
\eg singular value decomposition, we can hence decide both $\xx^*$
(and $\uu^* =\DD \xx^*$) and
$\bb^*$  up to a multiplicative scalar (note that we can never do
better with the information at hand, not even if we would be able to
solve \eqref{eq:probform}). The estimates of $\aa$ will be
identical for both problems (if the estimates are unique). 

The technique of introducing
the matrix $\X$ to avoid products between $\xx$ and $\bb$ is well
known in optimization and referred to as \textit{lifting} \citep{shor87,Lovasz91,Nesterov98,Goemans:1995}.

Problem \eqref{eq:probformm} is  a
non-convex optimization problem and  not  easier to solve than
\eqref{eq:probform}. To get an optimization problem we can solve, we
 remove the rank
constraint and instead minimize the rank. Since the rank of a matrix is
not a convex function, we replace the rank with a convex
heuristic. Here we choose the nuclear norm, but other heuristics are
also available (see for instance \cite{Fazel01arank}).  We then obtain the convex
 program   
\begin{subequations}\label{eq:probform3}
\begin{align}
\min_{\X, \aa}  \quad   \|\X\|_*& 
\\  \nonumber 
 \subjto \quad  y(t) -&\sum_{k_2=1}^{n_a} \a_{k_2}
 y(t-k_2) 
\\ \label{eq:probformm2} =& \sum_{k_1=1}^{n_b} (\DD \X)(t-n_k-k_1,k_1) , \;
 t=n,\dots,N,
\end{align}\end{subequations}
which we refer to as \textit{blind identification via
  lifting} (BIL).

Last, in the noisy setting we have to tolerate some nonzero modeling
error. If the noise $e$ is known to be bounded, say that $|e(t)|\leq \epsilon$, we suggest to use
\begin{subequations}\label{eq:probform332}
\begin{align}
\min_{\X, \aa, {\bf \eta}}  \quad    \|\X\|_*  
\\ \nonumber
 \subjto \quad  y(t) -&\sum_{k_2=1}^{n_a} \a_{k_2}
 y(t-k_2)
\\  \label{eq:probformm33} = & \sum_{k_1=1}^{n_b} (\DD
\X)(t-n_k-k_1,k_1)+\eta(t), \\
 | \eta(t) | \leq &\epsilon, \quad 
  t=  n,\dots,N,
\end{align}\end{subequations}
and if the noise is Gaussian,
\begin{subequations}\label{eq:probform32}
\begin{align}
\min_{\X, \aa, {\bf \eta}}  \quad    \|\X\|_*  + &\lambda \|{\bf \eta} \|_2^2 
\\ \nonumber
 \subjto \quad  y(t) -&\sum_{k_2=1}^{n_a} \a_{k_2}
 y(t-k_2)
\\  \label{eq:probformm3} = & \sum_{k_1=1}^{n_b} (\DD \X)(t-n_k-k_1,k_1)+\eta(t), \\
  t=&  n,\dots,N.
\end{align}\end{subequations}
In the latter case, we see $\lambda>0$ as a design parameter  and seek the largest
$\lambda $ such that $\X$ becomes rank 1.

\section{Analysis}
The number of optimization variables  in \eqref{eq:probform} is
essentially $n_a+n_b+m$, under the assumption that $\uu=\DD\xx$. We can
hence not expect a  reliable identification result from fewer than
$n_a+n_b+m$  measurements. One may wonder how many measurements that
are needed. Using that the constraint \eqref{eq:probformm2} of BIL is
linear in $\X$, we have the following result:
\begin{thm}[Guaranteed Recovery using BIL]\hfill

Consider the noise free blind ARX identification
problem~\eqref{eq:probform} and assume that it has a  unique solution
(up to a multiplicative scalar).
Let the row vector $d_i \in \Re^m$ be the $i$:th row of $\DD$. 
If  ${\bf A}=$
\begin{equation*}{\tiny
\left [\begin{smallmatrix}
d_{n-n_k-1} & d_{n-n_k-2} & \dots & d_{n-n_k-n_b} & y(n-1) & \dots & y(n-n_a)\\
d_{n-n_k}     & d_{n-n_k-1}        & \dots & d_{n-n_k-n_b+1} & y(n) & \dots & y(n-n_a+1)\\
\vdots &&\ddots &\vdots & \vdots &  & \vdots \\
d_{n-n_k-2+n_b } &\hdots &&d_{n-n_k-1} & y(n+n_b-2) & \dots & y(n-n_a+n_b-1)
\\
\vdots &&&\vdots & \vdots &  & \vdots
\\
d_{N-n_k-1} &d_{N-n_k-2}& \hdots &d_{N-n_k-n_b} & y(N-1) & \dots & y(N-n_a)
\end{smallmatrix} \right ]
}
\end{equation*}
has full column rank, then the ARX model and input solving
\eqref{eq:probform}  are recovered, up to
a multiplicative scalar, by BIL.
\end{thm}
\begin{pf}
From assumption, \eqref{eq:probform} has a unique solution. Form
$\X^*$ by multiplying the solutions, $\xx^*$ and $\bb^*$, of
\eqref{eq:probform}. That is,
$\X^*=\xx^* (\bb^*)^{\T}$. Let $a_1^*,\dots,a_{n_a}^*$ denote the
corresponding values for $a_1,\dots,a_{n_a}$ that solve
\eqref{eq:probform} and define $\theta^*$ as \begin{equation*}
\theta^*=\begin{bmatrix} \X^*(:,1)^{\T} & \X^*(:,2)^{\T} &\dots & \X^*(:,n_b)^{\T} &
  -a^*_1& \dots & -a^*_{n_a} \end{bmatrix}^{\T}.
\end{equation*} 
We must have that
 \begin{equation}\label{eq:uniq}
\begin{bmatrix} y(n) & y(n+1) & \dots & y(N) \end{bmatrix}^{\T}= {\bf A}
\theta^*. \end{equation}
Note that the pair $\X^*$ and $\aa^*$ is a feasible solution to \eqref{eq:probform3}.  Since the solution to \eqref{eq:uniq} must be  unique since ${\bf A}$
has full column rank, we must have that
BIL gives $\theta^*$. 
\qed
\end{pf}
Note that if the linear constraints of
\eqref{eq:probform3} alone give the solution of BIL, no optimization
is necessary. Seeking the
matrix $\X$ that gives the minimum nuclear norm is only of interest if
we have too few measurements for the constraints to uniquely define
the solution but more
measurements than
$n_a+n_b+m$.

The noisy case is harder to analyze and we leave the analysis as
future work.

\section{Computing $\lambda^{\text{\MakeLowercase{min}}}$}
In the noisy version \eqref{eq:probform32} of BIL, the design parameter $\lambda$ has to be
chosen. Since $\lambda$ regulates the tradeoff between the nuclear
norm and the squared norm of the estimated noise $\eta$, it is natural
to seek the largest $\lambda$ such that the estimate $\X$ is rank
1. In seeking this $\lambda$, the value for $\lambda
^{\text{\MakeLowercase{min}}}$ may come handy. $\lambda
^{\text{\MakeLowercase{min}}}$ is defined as the largest $\lambda$
such that $\X=0$ in BIL. Since the estimate for $\X$ will stay the
same for all $\lambda \leq \lambda
^{\text{\MakeLowercase{min}}}$, we should limit our search of
$\lambda$ to be within  $[\lambda
^{\text{\MakeLowercase{min}}} \, \infty]$. One may for example start
with $\lambda=\lambda
^{\text{\MakeLowercase{min}}} $ and then successively increase
$\lambda$ as long as $rank(\X)=1$.

\begin{thm}[Computing $\lambda_{\text{min}} $]\hfill

Consider the optimization
problem given in
\eqref{eq:probform32}. There exists a $\lambda$, denoted
$\lambda^{\text{min}} $, such that whenever $\lambda\leq
\lambda^{\text{min}} $, solving \eqref{eq:probform32} results in
$\X=0$. $\lambda^{\text{min}} $ is given by:
\begin{subequations}\label{eqlambdamax}
\begin{align} 1/\lambda^{\text{min}}  =\argmin_{{\bf V}
    \in\Re^{m\times n_b}}  \quad &
  \|{\bf V}\| \\ \nonumber \subjto \quad & 0=  {\bf V }(i,j) - 2 \sum_{t=n}^N\bigg(
y(t)  \\ \label{eq:sip4}-\sum_{k_2=1}^{n_a} &\hat \a_{k_2} 
 y(t-k_2) \bigg) 
   \DD(t-n_k-j,i), \\ & i=1,\dots, m, \,j=1,\dots,n_b,
\end{align}
\end{subequations}
with 
\begin{equation}
\{\hat a_1,\dots,\hat a_{n_a}\} = \argmin_{\aa}  \sum_{t=n}^N\bigg(
y(t)-\sum_{k_2=1}^{n_a} \a_{k_2}
 y(t-k_2) \bigg)^2.
\end{equation}
\end{thm}
\begin{pf}
The noisy version of BIL can be rewritten to take the form
\begin{align} \nonumber
\min_{\X, \aa}  &\quad    \|\X\|_*  + \lambda \sum_{t=n}^N\bigg(
y(t)\\ \label{eq:sip}-&\sum_{k_2=1}^{n_a} \a_{k_2}
 y(t-k_2)
-  \sum_{k_1=1}^{n_b} (\DD \X)(t-n_k-k_1,k_1) \bigg)^2.
\end{align}
The nuclear norm is not differentiable and it follows that for $\X=0$ to be a valid solution,
zero needs to
be in the subdifferential of the objective with respect to $\X$
evaluated at $\X=0$ (see \eg~\citet[Prop.~4.7.2]{Bert03}).
The subdifferential of the objective of \eqref{eq:sip} at $\X=0$ and
$\aa=\hat\aa$ with
respect to the $(i,j)$th element of $\X$ can be shown
equal to
\begin{align} 
       {\bf V }(i,j) - 2 \lambda \sum_{t=n}^N\bigg(
y(t)   \label{eq:sip45}-\sum_{k_2=1}^{n_a} &\hat \a_{k_2} 
 y(t-k_2) \bigg) 
   \DD(t-n_k-j,i).
\end{align}
 We further have that $\|{\bf V}\|\leq 1$ from the subdifferential of
 the nuclear norm (see for instance \citet{Watson199233} or \citet{recht10}). $\|\cdot \|$ is here  the operator norm (the largest singular
value). To find $\lambda_{\text{min}} $ we could now consider the
optimization problem 
\begin{subequations}
\begin{align} \max_{{\bf V}
    \in\Re^{m\times n_b} ,\lambda}  \quad &
 \lambda \\ \nonumber \subjto \quad & 0=  {\bf V }(i,j) - 2 \lambda \sum_{t=n}^N\bigg(
y(t)  \\ \label{eq:sip46}-\sum_{k_2=1}^{n_a} &\hat \a_{k_2} 
 y(t-k_2) \bigg) 
   \DD(t-n_k-j,i), \\ & i=1,\dots, m, \,j=1,\dots,n_b,\\\|{\bf V}&\|\leq 1,
\end{align}
\end{subequations}
which can be shown equivalent to \eqref{eqlambdamax}.
\qed
\end{pf}

$\lambda_{\text{min}} $ was also numerically verified. 

\section{Solution Algorithms and Software}

Many standard methods of convex optimization can be used to solve
problem \eqref{eq:probform3},
\eqref{eq:probform332} and \eqref{eq:probform32}. 
Systems such as CVX \citep{cvx1,cvx2} or YALMIP \citep{Yalmip}
can readily handle the nuclear norm. For large scale problems, the \textit{alternating direction
method of multipliers} (ADMM, see \eg \cite{bert:97,boyd:11}) is an attractive choice and we have
previously shown that ADMM can be very efficient on similar
problems \cite{ohlsson:13}. Code for solving
\eqref{eq:probform3}, \eqref{eq:probform332} and \eqref{eq:probform32} will
be made
available on \url{http://www.rt.isy.liu.se/~ohlsson/code.html}


\section{Numerical Illustration}
Consider the system given in the diagram below. 
\begin{center}
\tikzstyle{int}=[draw,line width=1pt, minimum size=1em]
\tikzstyle{init} = [pin edge={<-,black,line width=1pt},line width=1pt]
\begin{tikzpicture}[node distance=3.5cm,auto, >=latex',line width=1pt]
    \node [int] (a) {\small ZOH};
    \node (b) [left of=a,node distance=1cm, coordinate] {a};
    \node [int, pin={[init]above:$e$},line width=1pt] (c) [right of=a]
    {{\small $\begin{aligned}
  y(t)-& a_1y(t-1) =  b_1u(t-1) \\ +&   b_2 u(t-2)
          + b_3u(t-3) +e(t) \end{aligned}$ }};
    \node [coordinate] (end) [right of=c, node distance=3cm]{};
    \path[->] (b) edge node {$x$} (a);
    \path[->] (a) edge node {$u$} (c);
    \draw[->] (c) edge node {$y$} (end) ;
\end{tikzpicture}\end{center}
\bigskip

\begin{figure}
\includegraphics[width=0.99\columnwidth]{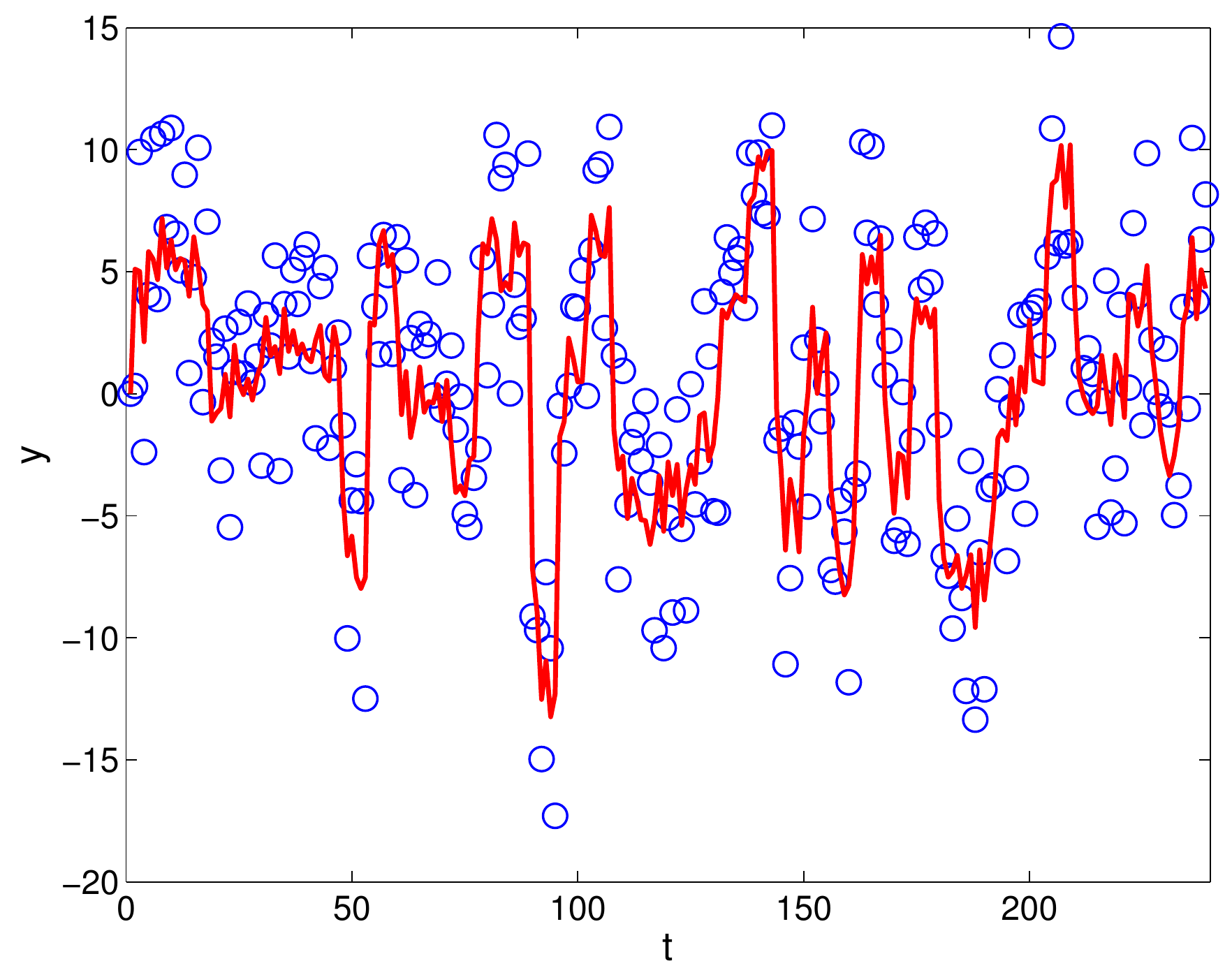}
\caption{The noise free (solid line) and noisy outputs (circles).}
\label{fig:ydata}
 \end{figure}

Here the values $x$ were generated by independently sampling from a unit Gaussian
and 
the noise $e$ by independently sampling from  a uniform distribution between
$-\epsilon/2$ and $\epsilon/2$. The
ZOH (zero-order hold) block holds the input to the ARX system constant for 6 consecutive
samples. We can therefore express $\uu$ in terms of $\xx$ as
\begin{equation}
\uu= \begin{bmatrix} \mathbbm{1}_{6\times 1} & \mathbb{0}_{6 \times
    1} &\mathbb{0}_{6\times 1} &\dots &\mathbb{0}_{6\times 1}  \\ \mathbb{0}_{6\times 1} & \mathbbm{1}_{6 \times
    1} &\mathbb{0}_{6\times 1} &\dots & \mathbb{0}_{6\times 1} \\ \vdots &&\ddots & &\vdots \\ \mathbb{0}_{6\times 1} &\dots &&\mathbbm{1}_{6 \times
    1}  & \mathbb{0}_{6 \times   1}
\\
  \mathbb{0}_{6\times 1} &\dots & &\mathbb{0}_{6\times 1} & \mathbbm{1}_{6 \times
    1} \end{bmatrix} \xx.
\end{equation}We identify the matrix in the relation between $\uu$ and
$\xx$ as $\DD$. 
The ARX coefficients used were  
\begin{equation} a_1=-0.3,\, b_3=1,\, b_2=2, \,b_1=3.\end{equation} 
Figure~\ref{fig:ydata}  shows the output $y$ for $\epsilon =5$. 

\begin{figure}
\includegraphics[width=0.99\columnwidth]{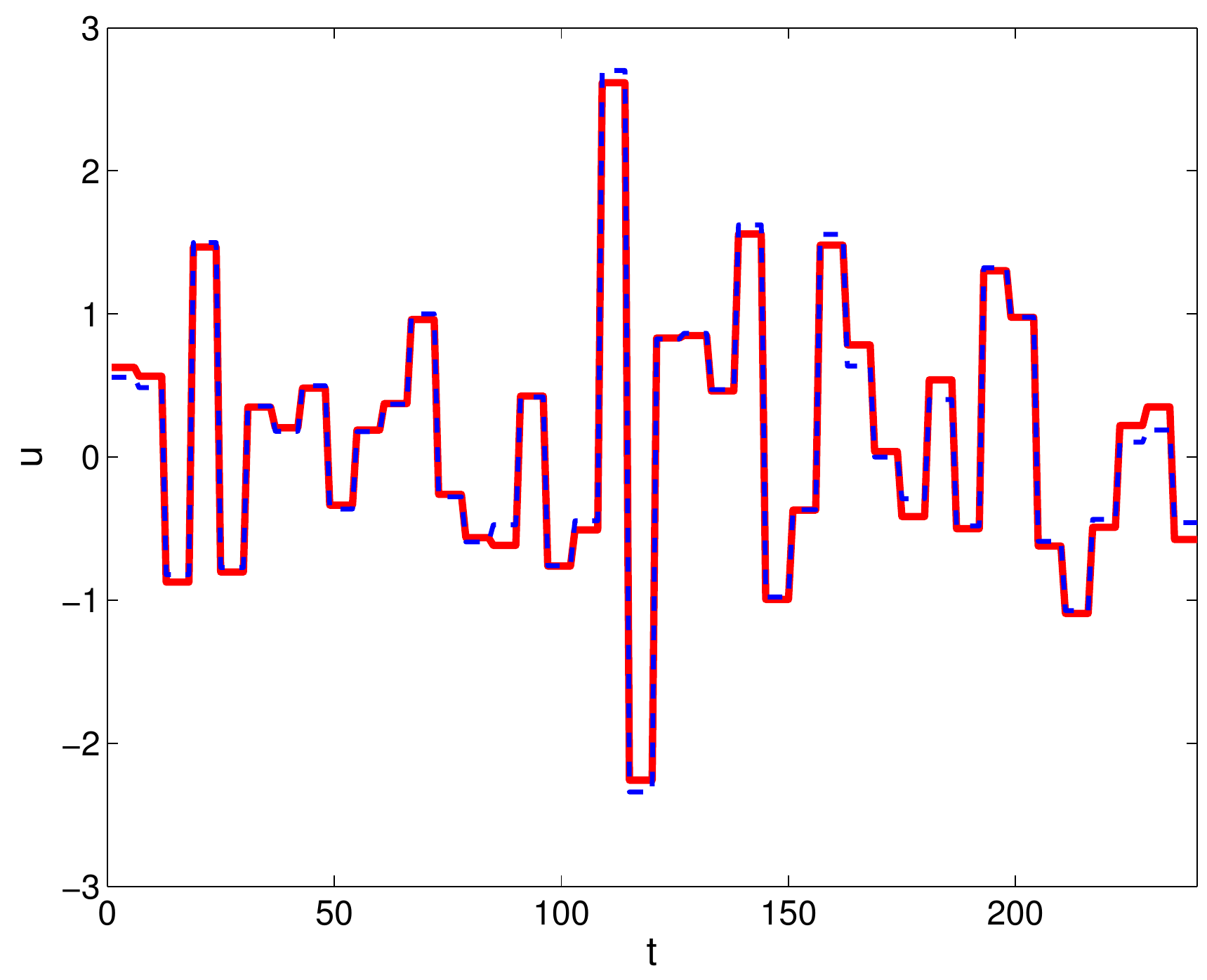}
\caption{The estimated (dashed line) and the true input $\uu$ (solid line).}
\label{fig:udata}
 \end{figure}

If the noisy version of  BIL \eqref{eq:probform332} is used to estimate $\uu$ and an ARX model, we get the input-estimate given in Figure~\ref{fig:udata} and the ARX
coefficients:
\begin{equation} \label{eq:bilest}
a_1=-0.21,\, b_3=0.91,\,   b_2=1.80,\,   b_1=2.7. 
\end{equation}
It is interesting to notice that if we instead would be given the true input
$\uu$ and only estimated the ARX coefficients by minimizing the squared
residuals between the output $\yy$ and the predicted output, we would
have got the estimates:   
\begin{equation} \label{eq:LSest}
a_1=-0.30,\, b_3=0.88,\,   b_2=2.39,\,   b_1=2.85. 
\end{equation}
As seen, these estimates are not that much better than what BIL is
providing (see \eqref{eq:bilest}). Remember that BIL is
only given the $y$-measurements and not the inputs $\uu$. It is
therefore quite remarkable that the estimates of BIL is comparable to
those given in \eqref{eq:LSest}. 

\begin{figure}
 \includegraphics[width=0.99\columnwidth]{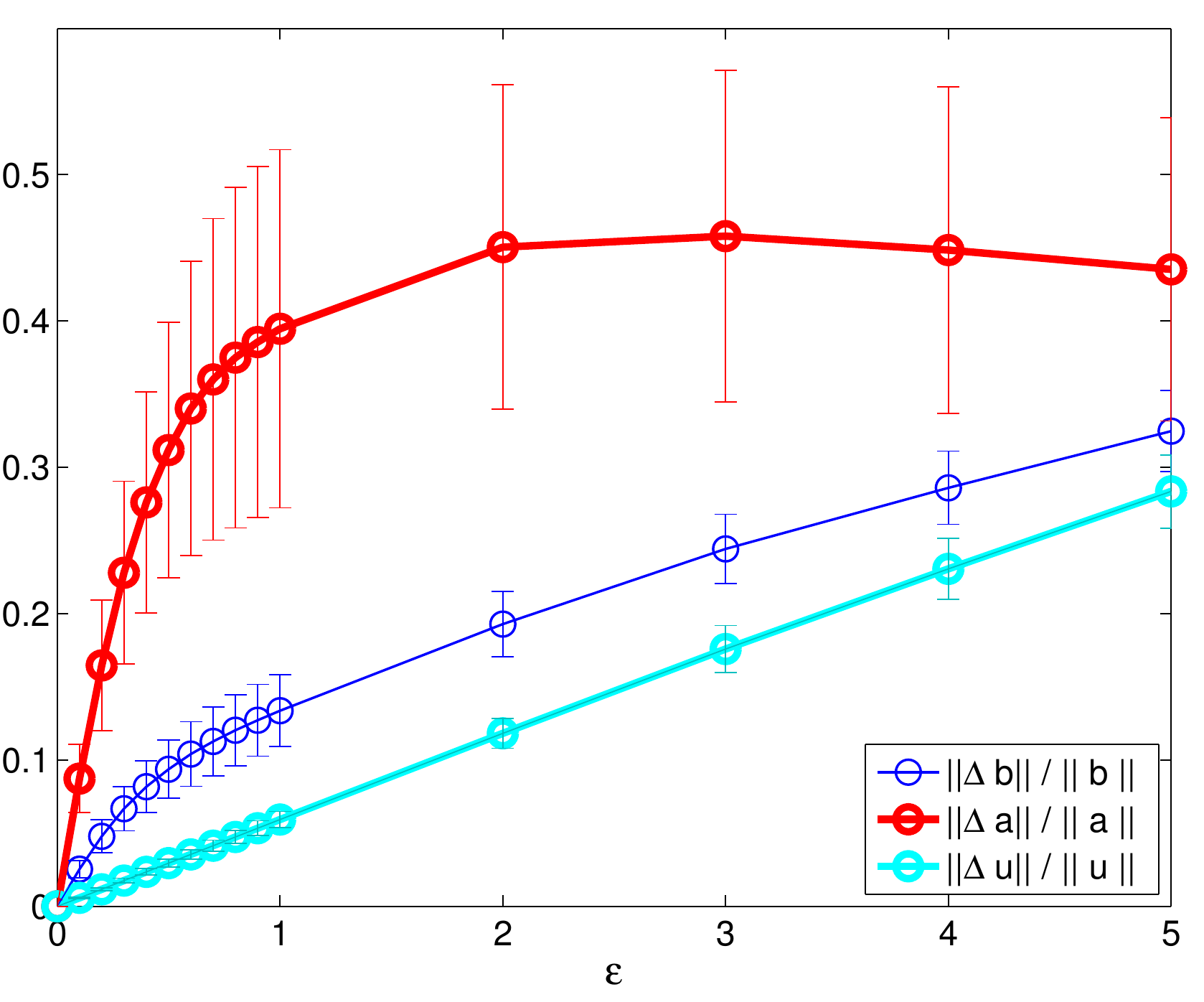}
\caption{The relative errors along with their 0.5 standard deviation
  error bounds for varying noise levels.}
\label{fig:error2}
 \end{figure}

To further study the robustness of BIL we carried out a Monte Carlo
simulation. In the simulation, the noise level $\epsilon$ was varied
between 0 and 5. For each noise level, 100 trials were carried out
with different noise and input realizations. The true ARX model was
kept fixed (the same as above). The results are summarized in
Figure~\ref{fig:error2}. 

The setup of above example does not give that ${\bf A}$ has full
column rank. Nevertheless, a perfect result was obtained in the noise free
case.  It can however be verified that if  $\DD$ is instead generated by
independently sampling each element from a unit Gaussian distribution
(everything else unchanged),
the resulting ${\bf A}$ has full column rank.

\section{Conclusion}
This paper presented a novel framework for blind system identification
of ARX model. The framework uses the fact that the problem can be
rewritten as a rank minimization problem. A convex relaxation is
presented to approximate the sought ARX parameters and the unknown
inputs.  

\bibliography{refHO}
\end{document}